\documentclass[11pt]{article}
\usepackage{amsfonts}
\usepackage{mathrsfs}
\usepackage{amsmath}
\usepackage{amssymb}
\usepackage{graphicx}
\usepackage{float}

\newtheorem{lemma}{Lemma}[section]
\newtheorem{theorem}{Theorem}[section]
\newtheorem{corollary}{Corollary}[section]
\newtheorem{definition}{Definition}[section]
\newtheorem{proposition}{Proposition}[section]

\def\blemma{\begin{lemma}\sl{}\def\elemma{\end{lemma}}}
\def\btheorem{\begin{theorem}\sl{}\def\etheorem{\end{theorem}}}

\def\bproposition{\begin{proposition}\sl{}\def\eproposition{\end{proposition}}}

\def\beqlb{\begin{eqnarray}}\def\eeqlb{\end{eqnarray}}
\def\beqnn{\begin{eqnarray*}}\def\eeqnn{\end{eqnarray*}}

\def\qed{\hfill$\Box$\medskip}

\def\<{\langle}\def\>{\rangle}

\begin{document}

\bigskip

\bigskip

\centerline{\LARGE\bf Option pricing under jump diffusion model}

\bigskip
\centerline{Qian Li\footnote{E-mail:\tt lqsx0510@163.com} \, Li Wang\footnote{\,The corresponding author.\quad E-mail: \tt wangli@mail.buct.edu.cn} }

\smallskip
\centerline{College of Mathematics and Physics}

\centerline{Beijing University of Chemical Technology,\,Beijing 100029, P.R. China}

%\smallskip
%\centerline{E-mail: \tt qianli@mail.buct.edu.cn}

%\bigskip
%\centerline{Li Wang \,\footnote{\,The corresponding author.\, E-mail: \tt wangli@mail.buct.edu.cn}}

%\smallskip
%\centerline{College of Mathematics and Physics}

%\centerline{Beijing University of Chemical Technology,\,Beijing 100029, P.R. China}

%\smallskip
%\centerline{E-mail: \tt wangli@mail.buct.edu.cn}

\bigskip\bigskip

{\narrower{\narrower

\centerline{\bf Abstract}

\bigskip

We provide an European option pricing formula written
in the form of an infinite series of Black–Scholes-type
terms under double Lévy jumps model, where both the interest rate and underlying price are driven by Lévy process.
The series solution converges with a radius of convergence, and it is complemented by some 
numerical experiments to demonstrate its speed of convergence.

\bigskip


\bigskip


\par}\par}

\bigskip\bigskip

\section{Introduction}

\setcounter{equation}{0}

Black and Scholes (1973) made a breakthrough by proposing an elegant model with the underlying price following a
geometric Brownian motion and deriving an analytical formula for European option prices.
Merton(1976) extended the model to the more-general case when the underlying stock returns are generated by a mixture of 
both continuous and jump processes.  However, there are gaps between those models and the market data.
One of their main drawbacks is the constant volatility assumption, another is the constant interest rate. 
Heston (1993) contributed a lot to the literature by incorporating the CIR (Cox–Ingersoll–Ross)
model to describe the volatility process and deriving a closed-form pricing formula for European options.
Incorporating stochastic interest rate into stock option pricing model is another line of extension.
At the earliest, Merton has discussed option pricing under stochastic interest rate, Amin and Jarrow (1992)
derived a closed form stock option pricing formula under Merton-type interest rate based on Heath, Jarrow and Morton (1992)
framework. Sattayatham and Pinkham (2013) proposed an alternative option pricing model, in which asset
prices follow a stochastic volatility Lévy model with stochastic interest rate driven by the Hull–White process.
He and Zhu (2017) derived a closed-form pricing formula for European options in the form of an infinite
series which is under the Heston model with the interest rate being another random
variable following the CIR model. 

In conclusion, all the above results, they either
assume the asset price or the interest rate has continuous path. So far as we know, there are few results on the both
jump case. In our paper, we use jump-extended Black-Scholes model to describe the 
underlying asset process and jump-extended CIR-model to describe the interest rate regarding the existence of discontinuous jumps in interest rates.
Inspired by the idea revealed in Merton(1976), we provide pricing formulas written
in the form of an infinite series of Black–Scholes-type terms
first for options on a single stock, then for options on basket of stocks.  Following a similar method used in He and Zhu (2017),
we show that Black–Scholes-type term can also be written in the form of an infinite series a radius of convergence, and it is complemented by some 
numerical experiments to demonstrate its speed of convergence.

The remainder of the paper is organized as follows. In section 2 we
give the models of stock price and the interest rate. The main
results and the corresponding proofs are presented in section 3 and section 4.
In section 5 we will give data simulations to demonstrate its speed of convergence.

%%%%%%%%%%%% (Section 2) %%%%%%%%%%%%%%

\section{The model}

\subsection{Interest rate model with Lévy jump}

\setcounter{equation}{0}

The classical term structure of interest rate models, such as the Vasicek (1977) model, the Cox et al. (1985) (CIR) model, the
Heath et al. (1992) model, and other extended models for pricing interest rate derivatives (see e.g., the Hull and White (1990)
model), all assume that processes of state variables (such as the short-term interest rate, or the long-term interest rate,
or others) can be modeled by pure diffusion process. This assumption is inconsistent with many empirical studies (see
e.g., \cite{HW10,J04,Ly99,PS13}) regarding the existence of discontinuous jumps in interest rates. Therefore, the jump-diffusion processes
are better model for the interest rate dynamics, since allowing jumps in
interest rate model can capture the properties of skewness and excess kurtosis commonly.

In this paper, we assume the term structure of the interest rate is modelled as follows, 
 \beqlb\label{2.1}
 	d r(t)=k(a-r(t)) d t+\sigma_r \sqrt{r(t)} d W(t)+d\left(\sum_{i=1}^{N(t)}X_i-\lambda C_X t\right)
\eeqlb
where $\{W(t):t\geq 0\}$ is Brownian motion,  $\{N(t):t\geq 0\}$ is a Poisson process with constant intensity rate $\lambda>0$, and the sequence $\{X_i:i\geq 1\}$ denotes the magnitudes
of jump, which are assumed to be i.i.d. random variables with distribution $f_r(x)$ on nonegative real numbers, $C_X=EX$. 
Let $N(dtdx)$ be the Poisson random measure corresponding to $\sum_{i=1}^{N(t)}X_i$, then it has intensity $\lambda f_r(x)dtdx$.
Moreover, $\{W(t):t\geq 0\}$ and $\{N(t):t\geq 0\}$ are independent.

\subsection{Asset price with Lévy jump }

\setcounter{equation}{0}

The underlying asset price $S(t)$ satisfies the following stochastic differential equation : 
 \beqnn
d S(t)=S(t-)\left[(r(t)-\lambda_1C_Y )d t+\sigma d W_1(t)+ (Y-1) d N_1(t)\right] 
\eeqnn
where $\sigma>0$, $\{W_1(t):t\geq 0\}$ is Brownian motion,  $\{N_1(t):t\geq 0\}$ is a Poisson process with constant intensity rate $\lambda_1>0$ and $Y$ is a nonnegative variable with distribution $f_Y(y)$ and $C_Y=EY-1$. According to the It\^{o}'s formula which will be given in the next section,
 \beqnn
S(t)=S(0) \exp \left\{\int_0^t r(s) d s+\sigma W_1(t)-\lambda_1 C_Yt-\frac{1}{2} \sigma^2 t\right\}\prod_{i=1}^{N_1(t)}Y_i  
\eeqnn
where $\{Y_i:i\geq 1\}$ are independent variables and has the same distribution as $Y$.
It's obvious that $S(t)$ is strictly positive when $Y_i>0$.

We further assume that all components of risk in the interest rate and asset prices dynamics,
including jump risks, are defined on a filtered complete probability space $(\Omega, \mathcal{F},Q)$
with information filtration $\{\mathcal{F}_t:t\geq 0\}$ satisfying the usual conditions, $\mathrm{Q}$ is the risk-neutral measure.
We assume $\{X_i:i\geq 1\}$, $Y$, $\{W(t):t\geq 0\}$, $\{N(t):t\geq 0\}$, $\{W_1(t):t\geq 0\}$ and $\{N_1(t):t\geq 0\}$ are independent.

\section{European option pricing}

\setcounter{equation}{0}

Under the risk-neutral measure $Q$, European call option $U(S, r, t)$ is actually the expectation of the present value of return,
$$
U(S,r,t)=E^Q\left[e^{-\int_t^T r(s) d s} \max (S(T)-K, 0) \mid S(t),r(t)\right] 
$$
where $K$ represents the strike price. Due to the existence of stochastic interest rate, there are two different random variables in the above formula, which is difficult to calculate. Therefore, in order to obtain the analytical expression of European call option price, we introduce forward measure $Q^{\top}$ (for the definition, see Brigo and Mercurio (2006)), then we can get
\beqlb
U(S,r,t)=b(t, T, r(t)) E^{Q^{\top}}[\max (S(T)-K, 0) \mid S(t),r(t)] 
\eeqlb
where $b(t, T, r(t))$ is the price of a zero-coupon bond with maturity $T$ under the risk-neutral measure $Q$,
which need to be detemined first.

%This means that we first have to calculate the formula of $b(t, T, r(t))$ (i.e. Theorem 4.1) and the dynamics of our model under $Q^{\top}$ (i.e. Corollary 3.1.2). 

\btheorem\label{t3.1} If the interest rate $r(t)$ is given by equation (\ref{2.1}), then the price $b(t, T, r(t))$ of a zero-coupon bond with maturity T at time t is
$$
b(t, T, r(t))=\exp \{A(t, T)+G(t, T) r(t)\}, \quad 0 \leq t \leq T,
$$
with $b(T, T, r(T))=1$, where
$$
\begin{aligned}
	G(t, T)=&-\frac{2\left[e^{m(T-t)}-1\right]}{2 m+(k+m)\left(e^{m(T-t)}-1\right)}, m=\sqrt{k^2+\sigma_r^2}, \\
	A(t, T)=&-(k a-\lambda C_X)\left\{\frac{4}{(m-k)(m+k)} \ln \left[\frac{2 m+(m+k)\left(e^{m(T-t)}-1\right)}{2 m}\right]+\frac{2}{k-m}(T-t)\right\}\\
	&~~~~~~~~~~~ +\int_t^T \lambda \int_{0}^{+\infty}\left[e^{G(s, T) x}-1\right]f_r(x)dxds.
\end{aligned}
$$\etheorem

{\it Proof.} Under the risk-neutral measure, the price of a zero-coupon bond with maturity $T$ at time $t$ is 

$$b(t, T, r(t))=E^Q\left[e^{-\int_t^T r(s) d s} \mid r(u),0\leq u\leq t\right],\quad \mbox{with}~~ b(T, T, r(T))=1.$$

Let  $B(t)=\exp \left\{\int_0^t r(s) d s\right\}$.
Then the discounted asset price $\left\{\frac{b(t, T, r(t))}{B(\mathrm{t})}: 0 \leq t \leq \mathrm{T}\right\}$ should be a $Q$-martingale (Brigo and Mercurio (2006)), we get the following differential equations,
\beqlb\label{3.2}
\left\{\begin{array}{c}
	\frac{\partial b(t, T, r(t))}{\partial t}+(k(a-r(t))-\lambda C_X) \frac{\partial b(t, T, r(t))}{\partial r}-r(t) b(t, T, r(t))+\frac{1}{2} \sigma_r^2 r(t) \frac{\partial^2 b(t, T, r(t))}{\partial r^2} \\
	+\lambda \int_{0}^{+\infty}\left(b(t, T, r(t)+x)-b(t, T, r(t))\right)f_r(x)dx=0, \\
	b(T, T, r(T))=1.
\end{array}\right.
\eeqlb
We know from Duffie (2003) that the price of the zero-coupon bond is an exponential affine function of the interest rate, or equivalently, $b(t, T, r(t))$ has the following form,
\beqlb\label{3.3}
b(t, T, r(t))=\exp \{A(t, T)+G(t, T) r(t)\} .
\eeqlb
Substituting (\ref{3.3}) into (\ref{3.2}), we can get 
\beqlb\label{3.4}
\left\{\begin{array}{c}
	\frac{d G(t,T)}{d t}-G(t,T) k-1+\frac{1}{2} G^2(t,T) \sigma_r^2=0, \\
	\frac{d A(t,T)}{d t}+G(t,T)( k a-\lambda C_X)+\lambda \int_{0}^{+\infty}\left[e^{G(t, T) x}-1\right]f_r(x)dx=0.
\end{array}\right.
\eeqlb
By solving (\ref{3.4}), we can get 
$$
G(t, T)=-\frac{2\left[e^{m(T-t)}-1\right]}{2 m+(k+m)\left(e^{m(T-t)}-1\right)}, m=\sqrt{k^2+\sigma_r^2},
$$
$$
\begin{aligned}
	A(t, T)=&-(k a-\lambda C_X)\left\{\frac{4}{(m-k)(m+k)} \ln \left[\frac{2 m+(m+k)\left(e^{m(T-t)}-1\right)}{2 m}\right]+\frac{2}{k-m}(T-t)\right\}\\
	&~~~~~~~~~~~ +\int_t^T \lambda \int_{0}^{+\infty}\left[e^{G(s, T) x}-1\right]f_r(x)dxds.
\end{aligned}
$$\qed

\blemma\label{l3.1} Sato (1999): Under measure $Q$, $\left\{W_t\right\}_{t \geq 0}$ is a standard Brownian motion, $N(dt,dx)$ is a Possion random measure with intensity $\lambda d t v(d x)$ defined on $[0, T] \times \mathbb{R}$, and $H(x)$ satisfies $\int_\mathbb{R} H(x) v (d x)<\infty$. There exists a martingale measure $\widetilde{Q}$ which is equivalent to measure $Q$, where $\frac{d\widetilde{Q}}{d Q} \mid \mathcal{F}_t=\xi_t$, and if
\beqnn
\xi_t=\exp \left\{\sigma W_t-\frac{1}{2} \sigma^2 t+\left(\int_\mathbb{R} \log H(x) N((0, t], d x)-t \int_\mathbb{R}(H(x)-1) v(d x)\right)\right\},
\eeqnn
then $\widetilde{W_t}=W_t-\sigma t$ is the standard Brownian motion and $\widetilde{N}(d t, d x)=N(d t, d x)-\lambda d t H(x) v(d x)$ is the Possion martingale under measure $\widetilde{Q}$.\elemma

\bproposition\label{p3.2} (Girsanov transform) The forward measure $Q^{\top}$ is defined by the following equation
$$
\begin{aligned}
	\frac{d Q^{\top}}{d Q}|\mathcal{F}_t=&\frac{b(t, T, r(t))}{b(0, T, r(0)) B(t)} \\
	=&\exp \left\{\int_0^t\left[G(s, T) \sigma_r \sqrt{r(s)} d W(s)-\frac{1}{2} G^2(s, T) \sigma_r^2 r(s) d s\right]\right.\\
	& +\left.\int_0^t \int_{0}^{+\infty} G(s, T) x N(d s, d x)-\int_0^t \int_{0}^{+\infty}\left(e^{G(s, T) x}-1\right)\lambda f_r(x)dx d s\right\}.
\end{aligned}
$$
Then under measure $Q^{\top}, \widetilde{W}(t):=W(t)-\int_0^t G(s, T) \sigma_r \sqrt{r(s)} d s$ is
a standard Brownian motion and $\widetilde{M}(d t, d x):=N(d t, d x)-e^{G(t, T) x} \lambda f_r(x)dx d t$
is a Possion martingale.\eproposition

{\it Proof.} By using It\^{o}'s formula and equation (\ref{3.4}), we obtain
$$
\frac{d b(t, T, r(t))}{b(t, T, r(t))}=r(t) d t+G(t, T) \sigma_r \sqrt{r(t)} d W(t)+\int_{0}^{+\infty}\left( e^{G(t, T) x}-
1 \right)\widetilde{N}(d t, d x)$$

therefore, 
$$
\begin{aligned}
&d \ln b(t, T, r(t))=\frac{1}{b(t, T, r(t))} d b^c(t, T, r(t))-\frac{1}{2 b^2(t, T, r(t))} d<b^c, b^c>\\
	&~~~~~~~+\int_{0}^{+\infty}\left\{\ln \left(b(t, T, r(t))+b(t, T, r(t))\left(e^{G(t, T) x}-1\right)\right)-\ln b(t, T, r(t))\right\} \widetilde{N}(d t, d x) \\
	&~~~~~~~+\int_{0}^{+\infty}\left\{\ln \left(b(t, T, r(t))+b(t, T, r(t))\left(e^{G(t, T) x}-1\right)\right)-\ln b(t, T, r(t))\right. \\
	&~~~~~~~~~~~~~~~~~~~~~~~~~~\left.-b(t, T, r(t))\left(e^{G(t, T) x}-1\right) \frac{1}{b(t, T, r(t))}\right\} \lambda f_r(x)dxdt\\
	&~~~~~~~=r(t) d t+G(t, T) \sigma_r \sqrt{r(t)} d W(t)-\frac{1}{2} G^2(t, T) \sigma_r^2 r(t) dt + G(t, T)\int_{0}^{+\infty} x \widetilde{N}(d t, dx) \\
	&~~~~~~~~~~~~~~~~~~~~~~~~~~-\lambda\int_{0}^{+\infty}\left(e^{G(t, T) x}-G(t, T) x-1\right) f_r(x)dx d t.
\end{aligned}
$$

where $d b^c(t, T, r(t)):=b(t, T, r(t))\left[r(t) d t+G(t,T) \sigma_r \sqrt{r(t)} d W(t)\right]$. Then we can get
$$
\begin{aligned}
	\frac{b(t, T, r(t))}{b(0, T, r(0))}=& \exp \left\{\int_0^t\left[r(s) d s+G(s,T) \sigma_r \sqrt{r(s)} d W(s)-\frac{1}{2} G^2(s,T) \sigma_r^2 r(s) d s\right]+\right.\\
	&\left.\int_0^t\int_{0}^{+\infty} G(s, T) x N(d s, dx)-\int_0^t \int_{0}^{+\infty}\left(e^{G(s, T) x}-1\right) \lambda f_r(x) dxd s\right\} .
\end{aligned}
$$
\qed

Under risk-neutral measure $\mathrm{Q}$, if the interest rate satisfies (\ref{2.1}), then under the forward measure $Q^{\top}$, 
$$
\begin{aligned}
  d r(t)&=k(a-r(t)) d t+\sigma_r \sqrt{r(t)} d \widetilde{W}(t)+G(t,T) \sigma_r^2 r(t) d t \\
 	      &~~~~+\int_{0}^{+\infty} x\left[\widetilde{M}(d t, d x)+\left(e^{G(t, T) x}-1\right) \lambda f_r(x)\right] d tdx\\
 	      &=\mu_r(t)dt+\sigma_r \sqrt{r(t)} d \widetilde{W}(t)+\int_{0}^{+\infty} x\widetilde{M}(d t, d x)
\end{aligned}
$$
where 
\beqlb\label{3.5}
\mu_r(t):=k(a-r(t))+G(t,T) \sigma_r^2 r(t)+\lambda\int_{0}^{+\infty}x\left(e^{G(t, T) x}-1\right)f_r(x) dx
\eeqlb
and the price of the underlying asset remains the same
$$d S(t)=S(t-)\left[(r(t)-\lambda_1C_Y )d t+\sigma d W_1(t)+ (Y-1) d N_1(t)\right]. $$

\noindent In the next section, based on the results obtained in this section, the explicit pricing formula of European options is derived by calculating the expectation of the payment function under the measure $Q^{\top}$.

\subsection{Derivation of pricing formula}
	
Under the forward measure $Q^{\top}$, the price of a call option with maturity $\mathrm{T}$ and strike price $\mathrm{K}$ can be obtained as follows:
$$
U(S, r, t)=b(t, T, r(t)) E^{Q^{\top}}\left[\max (S(T)-K, 0) \mid S(t), r(t)\right]
$$
where $b(t, T, r(t))$ has been given by Theorem \ref{t3.1}.

Let $F\left(S, r, \tau\right)=E^{Q^{\top}}\left[\max (S(T)-K, 0) \mid S(t), r(t)\right]$, where $\tau=T-t$. According to the Feynman-Kac
formula,
\beqlb\label{3.6}
\begin{gathered}
	\left(r-\lambda_1 C_Y\right) S F_S+\frac{1}{2} \sigma^2 S^2 F_{S S}+\mu_rF_r	+\frac{1}{2} \sigma_r^2 r(t) F_{r r}-F_\tau\\
		+\lambda \int_{0}^{\infty}\left(F\left(S, r+x, \tau\right)-F\left(S, r, \tau\right)\right) f_r(x)dx\\
+\lambda_1 \int_{0}^{\infty}\left(F\left(S y, r, \tau\right)-F\left(S, r, \tau\right)\right) f_{Y}(y)dy=0 .
\end{gathered} 
\eeqlb

\noindent In order to solve equation (\ref{3.6}), we introduce the following equations,
\beqlb\label{3.7}
\left\{\begin{array}{c}
	d S^c(t)=S^c(t)\left[r^c(t) d t+\sigma d W_1(t)\right] \\
	d r^c(t)=\mu_{r^c}(t) d t+\sigma_r \sqrt{r^c(t)} d \widetilde{W}(t)
\end{array}\right.
\eeqlb
where $\mu_{r^c}(t)$ has been defined by (\ref{3.5}) with $r$ replaced by $r^c$. Let $W(x,r,t)$ be the function defined by 
\beqnn
W\left(S, r, \tau\right)=E^{Q^{\top}}\left[\max \left(S^c(T)-K, 0\right) \mid S^c(t), r^c(t)\right]|_{S^c(t)=S,r^c(t)=r}, ~\tau=T-t. 
\eeqnn
According to the Feynman-Kac formula,
\beqlb\label{3.8}
r S W_{x}+\frac{1}{2} \sigma^2 S^2 W_{x x}+\mu_{r}W_{r}+\frac{1}{2} \sigma_r^2 r W_{r r}-W_\tau=0.
\eeqlb

\btheorem\label{t3.2}
The solution to (\ref{3.6}) is
\beqlb\label{3.9}
\begin{gathered}
	F\left(S, r, \tau\right)=\sum_{l=0}^{\infty}\sum_{n=0}^{\infty} \frac{\left(\lambda \tau\right)^l}{l !} e^{-\lambda\tau} \frac{\left(\lambda_1\tau\right)^n}{n !} e^{-\lambda_1 \tau} 
	E_{l, n}\left[W\left(S\left(\prod_{i=1}^n Y_i\right) e^{-\lambda_1C_Y\tau}, r+\sum_{i=1}^lX_i, \tau\right)\right]
\end{gathered} 
\eeqlb
where ''$E_{l, n}$" is the expectation over the distribution of $\sum_{i=1}^lX_i$ and  $\prod_{i=1}^n Y_i$.
\etheorem

{\it Proof.} It only needs to verify that (\ref{3.9}) satisfies Equation (\ref{3.6}).
Let 
$$
\begin{gathered}
    V_n:=S\left(\prod_{i=1}^nY_i\right) e^{-\lambda_1C_Y\tau}, \quad
	P_l(\tau):=\frac{\left(\lambda \tau\right)^l}{l !} e^{-\lambda \tau}, \quad P_n^1(\tau):=\frac{\left(\lambda_1 \tau\right)^n}{n !} e^{-\lambda_1 \tau}. 
\end{gathered}
$$

So,
$$
\begin{aligned}
	S F_S &=\sum_{l=0}^{\infty} \sum_{n=0}^{\infty} P_l(\tau) P_n^1(\tau) E_{l,n}\left(V_{n} W_x\right) \\
	S^2 F_{S S} &=\sum_{l=0}^{\infty} \sum_{n=0}^{\infty} P_l(\tau) P_n^1(\tau) E_{l,n}\left(V_{n}^2 W_{xx}\right)
\end{aligned}
$$
$$
\mu_{r}(t) F_r=\sum_{l=0}^{\infty} \sum_{n=0}^{\infty} P_l(\tau) P_n^1(\tau) E_{l,n}\left(W_r \cdot \mu_{r}(t)\right)
$$
$$
\sigma_r^2 r(t) F_{r r}=\sum_{l=0}^{\infty} \sum_{n=0}^{\infty} P_l(\tau) P_n^1(\tau) E_{l,n}\left(W_{r r} \cdot \sigma_r^2 \cdot r(t)\right)
$$
$$
\begin{aligned}
F_\tau=&-\lambda F-\lambda_1F-\lambda_1 C_Y \sum_{l=0}^{\infty} \sum_{n=0}^{\infty} P_l(\tau) P_n^1(\tau) E_{l,n}\left(V_{n} W_x\right)\\
&+\sum_{l=0}^{\infty} \sum_{n=0}^{\infty} P_l(\tau) P_n^1(\tau) E_{l,n}\left(W_\tau\right)\\
&+\lambda_1\sum_{l=0}^{\infty} \sum_{n=0}^{\infty} P_l(\tau) P_n^1(\tau) E_{l,n+1} \left[W\left(V_{n+1}, r+\sum_{i=1}^{l}X_i, \tau\right)\right]\\
&+\lambda\sum_{l=0}^{\infty} \sum_{n=0}^{\infty} P_l(\tau) P_n^1(\tau) E_{l+1,n}\left[W\left(V_{n}, r+\sum_{i=1}^{l+1}X_i, \tau\right)\right]
\end{aligned}
$$
Substitute all the above into (\ref{3.6}), then by using (\ref{3.8}), we get
$$
\begin{aligned}
  &\left(r-\lambda_1 C_Y\right) S F_S+\frac{1}{2} \sigma^2 S^2 F_{S S}+\mu_rF_r	+\frac{1}{2} \sigma_r^2 r(t) F_{r r}-F_\tau\\
	&=\sum_{l=0}^{\infty} \sum_{n=0}^{\infty} P_l(\tau) P_n^1(\tau) E_{l,n}\left(r V_{n} W_x\right)\\
	&~~+\frac{1}{2} \sigma^2 \sum_{l=0}^{\infty} \sum_{n=0}^{\infty} P_l(\tau) P_n^1(\tau) E_{l,n}\left(V_{n}^2 W_{xx}\right)\\
	&~~+\sum_{l=0}^{\infty} \sum_{n=0}^{\infty} P_l(\tau) P_n^1(\tau) E_{l,n}\left(W_r\cdot \mu_{r}(t)\right)\\
	&~~+\frac{1}{2} \sum_{l=0}^{\infty} \sum_{n=0}^{\infty} P_l(\tau) P_n^1(\tau) E_{l,n}\left(W_{r r} \cdot \sigma_r^2 r(t)\right)+\lambda_1 F+\lambda F\\
	&~~-\sum_{l=0}^{\infty} \sum_{n=0}^{\infty} P_l(\tau) P_n^1(\tau) E_{l,n}\left(W_\tau\right)\\
	&~~-\lambda_1 \sum_{l=0}^{\infty} \sum_{n=0}^{\infty} P_l(\tau) P_n^1(\tau) E_{l,n+1} \left[W\left(V_{n+1}, r+\sum_{i=1}^{l}X_i, \tau\right)\right]\\
	&~~-\lambda\sum_{l=0}^{\infty} \sum_{n=0}^{\infty} P_l(\tau) P_n^1(\tau) E_{l+1,n}\left[W\left(V_{n}, r+\sum_{i=1}^{l+1}X_i, \tau\right)\right]\\
	&=-\lambda \int_{0}^{\infty}\left(F\left(S, r+x, \tau\right)-F\left(S, r, \tau\right)\right) f_r(x)dx\\
	&~~~-\lambda_1 \int_{0}^{\infty}\left(F\left(S y, r, \tau\right)-F\left(S, r, \tau\right)\right) f_Y(y)dy.
\end{aligned}
$$
The desired result follows.
\qed

\section{Options on basket of stocks}

\setcounter{equation}{0}

Next, we consider options on multi-assets $\{S_i:1\leq i\leq n\}$, $n\geq 1$. 
The price of the call option with maturity $T$ and strike price $K$ has payoff $\max(H(S_1,\cdots,S_n)(T)-K,0)$.
For example, we may take $H(S_1,\cdots,S_n)=\sum_{i=1}^n \alpha_i S_i$ or  $\prod_{i=1}^nS^{\alpha_i}_i$, with $\alpha_i\geq 0$ and $\sum_{i=1}^n\alpha_i=1$.
For simplicity, we may assume $n=2$.

Without loss of generality, we assume that the interest rate $r(t)$ satisfies equation (\ref{2.1}) and
the stock prices satisfy the following equations,
$$
\begin{aligned}
d S_1(t)&=S_1(t-)\left[(r(t)-\lambda_1C_{Y_1} )d t+\sigma_1 d W_1(t)+ (Y_1-1) d N_1(t)\right]\\
d S_2(t)&=S_2(t-)\left[(r(t)-\lambda_2C_{Y_2} )d t+\sigma_2 d W_2(t)+ (Y_2-1) d N_2(t)\right]
\end{aligned}
$$
where $\sigma_1>0, \sigma_2>0$,  $Y_1$ and $Y_2$  are independent  nonnegative variables with distribution $f_{Y_1}(y)$ and $f_{Y_2}(y)$, respectively,
and $C_{Y_i}=EY_i-1$, $i=1,2$.
$\{W_1(t):t\geq 0\}$ and $\{W_2(t):t\geq 0\}$ are two correlated Brownian motions with correlation $d W_1(t) d W_2(t)=\rho dt$,
while $N_1(t)$ and $N_2(t)$ are two independent Poisson processes with intensity rates $\lambda_1>0$ and $\lambda_2>0$, respectively.
%Then
%$$
%\begin{aligned}
%d S_1(t)S_2(t)&=S_1(t-)dS_2(t)+S_2(t-)dS_1(t)+\rho S_1(t-)S_2(t-)\sigma_1\sigma_2 dt\\
 %          &=S_1(t-)S_2(t-)\left[(r(t)-\lambda_2C_{Y_2} )d t+\sigma_2 d W_2(t)+ (Y_2-1) d N_2(t)\right.\\
 %       &~~~~+\left.(r(t)-\lambda_1C_{Y_1} )d t+\sigma_1 d W_1(t)+ (Y_1-1) d N_1(t)\right]+\rho\sigma_1\sigma_2 dt.
%\end{aligned}$$
Under the forward measure $Q^{\top}$, the price of a call option with maturity $T$ and strike price $K$ can be obtained as follows,
\beqlb\label{4.1}
U(S, r, t)=b(t, T, r(t)) E^{Q^{\top}}\left[\max (H(S_1,S_2)(T)-K, 0) \mid S_1(t), S_2(t), r(t)\right]
\eeqlb
where $b(t, T, r(t))$ has been given by Theorem \ref{t3.1}. Let $G(x,y, r, \tau)$ be the function defined by
\beqnn
G(S_1,S_2, r, \tau):= E^{Q^{\top}}\left[\max (H(S_1,S_2)(T)-K, 0) \mid S_1(t), S_2(t), r(t)\right]
\eeqnn
where $\tau=T-t$. According to the Feynman-Kac formula,
\beqlb\label{4.2}
&&G_x(S_1,S_2,r,\tau)S_1(t)(r(t)-\lambda_1C_{Y_1})+G_y(S_1,S_2,r,\tau)S_2(t)(r(t)-\lambda_2C_{Y_2})\nonumber\\
&&~~~+G_{xy}(S_1,S_2,r,\tau)(t)\sigma_1\sigma_2S_1S_2\rho\nonumber \\
&&~~~+\frac{1}{2}G_{xx}(S_1,S_2,r,\tau)\sigma^2_1S^2_1 +\frac{1}{2}G_{yy}(S_1,S_2,r,\tau)\sigma^2_2S^2_2\nonumber \\
&&~~~+\mu_rG_r	+\frac{1}{2} \sigma_r^2 r G_{r r}-G_\tau\nonumber\\
&&~~~+\lambda \int_{0}^{\infty}\left(G\left(S_1,S_2, r+x, \tau\right)-G\left(S_1,S_2, r, \tau\right)\right) f_r(x)dx\nonumber \\
&&~~~+\lambda_1 \int_{0}^{\infty}\left(G\left(S_1y,S_2, r, \tau\right)-G\left(S_1,S_2, r, \tau\right)\right)f_{Y_1}(y)dy\nonumber \\
&&~~~+\lambda_2 \int_{0}^{\infty}\left(G\left(S_1,S_2y,, r, \tau\right)-G\left(S_1,S_2, r, \tau\right)\right)f_{Y_2}(y)dy=0.
\eeqlb

\noindent In order to solve equation (\ref{4.2}), we introduce the following equations,
\beqnn
\left\{\begin{array}{c}
d S_1^c(t)=S_1^c(t)\left[r^c(t) d t+\sigma_1 d W_1(t)\right] \\
d S_2^c(t)=S_2^c(t)\left[r^c(t) d t+\sigma_2 d W_2(t)\right] \\
d r^c(t)=\mu_{r^c}(t) d t+\sigma_r \sqrt{r^c(t)} d \widetilde{W}(t)
\end{array}\right.
\eeqnn
where $\mu_{r^c}(t)$ has been defined by (\ref{3.5}) with $r$ replaced by $r^c$. Let $W^H\left(x,y, r, \tau\right)$ be the function defined by
\beqnn
W^H\left(S_1,S_2, r, \tau\right)=E^{Q^{\top}}\left[\max \left(H(S_1^c(T),S_2^c(T))-K, 0\right) \mid S^c_1(t),S^c_2(t),r(t)\right]|_{S^c_1(t)=S_1,S^c_2(t)=S_2,r(t)=r}.
\eeqnn
 According to the Feynman-Kac formula,
\beqlb\label{4.3}
&W^H_x(S_1,S_2,r,\tau)S_1r+W^H_y(S_1,S_2,r,\tau)S_2r+W^H_{xy}(S_1,S_2,r,\tau)\sigma_1S_1\sigma_2S_2\rho \nonumber\\
&~~~+\frac{1}{2}W^H_{xx}(S_1,S_2,r,\tau)\sigma^2_1S^2_1 +\frac{1}{2}W^H_{yy}(S_1,S_2,r,\tau)\sigma^2_2S^2_2 \nonumber \\
&~~~+\mu_{r}W^H_r+\frac{1}{2} \sigma_r^2 r(t) W^H_{r r}-W^H_\tau=0.
\eeqlb

\btheorem\label{t4.1} The solution to (\ref{4.2}) is 
\beqlb\label{4.4}
&G\left(S_1,S_2, r, \tau\right)=\sum_{l=0}^{\infty}\sum_{n=0}^{\infty} \sum_{m=0}^{\infty}\frac{\left(\lambda \tau\right)^l}{l !}e^{-\lambda \tau} \frac{\left(\lambda_1 \tau\right)^n}{n !} e^{-\lambda_1 \tau} \frac{\left(\lambda_2 \tau\right)^m}{m !} e^{-\lambda_2 \tau} \nonumber\\
&~~~~~~~~~~~~\cdot E_{l,n, m}\left[W^H\left(S_1\prod_{i=1}^nY_{1i}e^{-\lambda_1C_{Y_1}\tau},S_2\prod_{j=1}^m Y_{2j} e^{-\lambda_2C_{Y_2}\tau}, r+\sum_{k=1}^lX_i, \tau\right)\right]
\eeqlb
where $Y_{1i}, i\geq 1$  ($Y_{2j}, j\geq 1$) are independent and identically distributed variables which have the same distribution as $Y_1$ ($Y_2$), and
''$E_{l, n,m}$" is the expectation over the distribution of $\sum_{i=1}^lX_i$, $\prod_{i=1}^n Y_{1i}$ and $\prod_{j=1}^m Y_{2j}$.
\etheorem

{\it Proof.} It only needs to verify that (\ref{4.4}) satisfies Equation (\ref{4.2}).
Let 
$$
\begin{gathered}
V_{n}:=S_1\prod_{i=1}^nY_{1i}e^{-\lambda_1C_{Y_{1}}\tau},U_{m}:=S_2\prod_{j=1}^m Y_{2j} e^{-\lambda_2C_{Y_2}\tau},n,m\geq 1\\
%R_l:=r+\sum_{i=1}^l X_i, l\geq 1,\\
P_l(\tau):=\frac{\left(\lambda\tau\right)^l}{l !} e^{-\lambda \tau}, \quad P_n^1(\tau):=\frac{\left(\lambda_1 \tau\right)^n}{n !} e^{-\lambda_1 \tau}, \quad P_m^2(\tau):=\frac{\left(\lambda_2 \tau\right)^m}{m !} e^{-\lambda_2 \tau}. \\
\end{gathered}
$$

So,
$$
\begin{aligned}
S_1 G_{x} &=\sum_{l=0}^{\infty}\sum_{n=0}^{\infty} \sum_{m=0}^{\infty}P_l(\tau) P_n^1(\tau) P_m^2(\tau) E_{l,n, m}\left(V_{n} W^H_x\right) \\
S_1^2 G_{xx} &=\sum_{l=0}^{\infty}\sum_{n=0}^{\infty} \sum_{m=0}^{\infty}P_l(\tau)P_n^1(\tau) P_m^2(\tau) E_{l,n, m}\left(V_{n}^2 W^H_{xx}\right)\\
S_2 G_{y} &=\sum_{l=0}^{\infty}\sum_{n=0}^{\infty} \sum_{m=0}^{\infty}P_l(\tau) P_n^1(\tau) P_m^2(\tau) E_{l,n, m}\left(U_{m} W^H_y\right) \\
S_2^2 G_{yy} &=\sum_{l=0}^{\infty}\sum_{n=0}^{\infty} \sum_{m=0}^{\infty}P_l(\tau)P_n^1(\tau) P_m^2(\tau) E_{l,n, m}\left(U_{m}^2 W^H_{yy}\right)\\
S_1S_2 G_{xy} &=\sum_{l=0}^{\infty}\sum_{n=0}^{\infty} \sum_{m=0}^{\infty}P_l(\tau)P_n^1(\tau) P_m^2(\tau) E_{l,n, m}\left(V_nU_m W^H_{xy}\right)
\end{aligned}
$$
$$
\mu_r(t) G_r=\sum_{l=0}^{\infty}\sum_{n=0}^{\infty} \sum_{m=0}^{\infty}P_l(\tau) P_n^1(\tau) P_m^2(\tau) E_{l,n, m}\left(W^H_r \mu_r(t)\right)
$$
$$
\sigma_r^2 r(t) G_{r r}=\sum_{l=0}^{\infty}\sum_{n=0}^{\infty} \sum_{m=0}^{\infty}P_l(\tau) P_n^1(\tau) P_m^2(\tau) E_{l,n, m}\left(W^H_{r r} \sigma_r^2 r(t)\right)
$$
$$
\begin{aligned}
G_\tau&=-\lambda G-\lambda_1 G-\lambda_2G-\lambda_1 C_{Y_1}\sum_{l=0}^{\infty}\sum_{n=0}^{\infty} \sum_{m=0}^{\infty}P_l(\tau) P_n^1(\tau) P_m^2(\tau) E_{l,n, m}\left(V_{n} W^H_x\right)\\
&~~~-\lambda_2 C_{Y_2}\sum_{l=0}^{\infty}\sum_{n=0}^{\infty} \sum_{m=0}^{\infty}P_l(\tau) P_n^1(\tau) P_m^2(\tau) E_{l,n, m}\left(U_{m} W^H_y\right)\\
&~~~+\sum_{l=0}^{\infty}\sum_{n=0}^{\infty} \sum_{m=0}^{\infty}P_l(\tau) P_n^1(\tau) P_m^2(\tau) E_{n, m}\left(W^H_\tau\right)\\
&~~~+\lambda \sum_{l=0}^{\infty}\sum_{n=0}^{\infty} \sum_{m=0}^{\infty}P_l(\tau)P_n^1(\tau) P_m^2(\tau) E_{l+1,n, m}\left[W^H\left(V_{n,m}, r+\sum_{i=1}^{l+1}X_i, \tau\right)\right]\\
&~~~+\lambda_1 \sum_{l=0}^{\infty}\sum_{n=0}^{\infty} \sum_{m=0}^{\infty}P_l(\tau)P_n^1(\tau) P_m^2(\tau) E_{l,n+1, m}\left[W^H\left(V_{n+1,m}, r+\sum_{i=1}^{l}X_i, \tau\right)\right]\\
&~~~+\lambda_2\sum_{l=0}^{\infty}\sum_{n=0}^{\infty} \sum_{m=0}^{\infty}P_l(\tau)P_n^1(\tau) P_m^2(\tau) E_{l,n, m+1}\left[W^H\left(V_{n,m+1}, r+\sum_{i=1}^{l}X_i, \tau\right)\right]
\end{aligned}
$$
Substitute all the above into (\ref{4.2}) and use (\ref{4.3}), we get
$$
\begin{aligned}
	&G_x(S_1,S_2,r,\tau)S_1(t)(r(t)-\lambda_1C_{Y_1})+G_y(S_1,S_2,r,\tau)S_2(t)(r(t)-\lambda_2C_{Y_2})\nonumber\\
	&~~+G_{xy}(S_1,S_2,r,\tau)(t)\sigma_1\sigma_2S_1S_2\rho\nonumber \\
	&~~+\frac{1}{2}G_{xx}(S_1,S_2,r,\tau)\sigma^2_1S^2_1 +\frac{1}{2}G_{yy}(S_1,S_2,r,\tau)\sigma^2_2S^2_2\nonumber \\
	&~~+\mu_rG_r	+\frac{1}{2} \sigma_r^2 r G_{r r}-G_\tau\nonumber\\
	&=\sum_{l=0}^{\infty}\sum_{n=0}^{\infty} \sum_{m=0}^{\infty}P_l(\tau) P_n^1(\tau) P_m^2(\tau) E_{l,n, m}\left(r V_{n} W^H_x\right)\\
	&~~~+\sum_{l=0}^{\infty}\sum_{n=0}^{\infty} \sum_{m=0}^{\infty}P_l(\tau) P_n^1(\tau) P_m^2(\tau) E_{l,n, m}\left(r U_{m} W^H_y\right)\\
	&~~~+\frac{1}{2} \sigma_1^2\sum_{l=0}^{\infty} \sum_{n=0}^{\infty} \sum_{m=0}^{\infty}P_l(\tau) P_n^1(\tau) P_m^2(\tau) E_{l,n, m}\left(V_{n}^2 W^H_{xx}\right)\\
	&~~~+\frac{1}{2} \sigma_2^2\sum_{l=0}^{\infty} \sum_{n=0}^{\infty} \sum_{m=0}^{\infty}P_l(\tau) P_n^1(\tau) P_m^2(\tau) E_{l,n, m}\left(U_{m}^2 W^H_{yy}\right)\\
	&~~~+\rho\sigma_1\sigma_2\sum_{l=0}^{\infty} \sum_{n=0}^{\infty} \sum_{m=0}^{\infty}P_l(\tau) P_n^1(\tau) P_m^2(\tau) E_{l,n, m}\left(V_{n}U_{m} W^H_{xy}\right)\\
	&~~~+\sum_{l=0}^{\infty} \sum_{n=0}^{\infty} \sum_{m=0}^{\infty}P_l(\tau) P_n^1(\tau) P_m^2(\tau) E_{n, m}\left(W^H_r\cdot \mu_r(t)\right)\\
	&~~~+\frac{1}{2} \sum_{l=0}^{\infty} \sum_{n=0}^{\infty} \sum_{m=0}^{\infty}P_l(\tau) P_n^1(\tau) P_m^2(\tau) E_{n, m}\left(W^H_{r r} \cdot \sigma_r^2 r(t)\right)+\lambda G+\lambda_1 G+\lambda_2 G\\
	&~~~-\sum_{l=0}^{\infty} \sum_{n=0}^{\infty} \sum_{m=0}^{\infty}P_l(\tau)P_n^1(\tau) P_m^2(\tau) E_{n, m}\left(W^H_\tau\right)
\end{aligned}
$$

$$
\begin{aligned}
&~~~-\lambda \sum_{l=0}^{\infty}\sum_{n=0}^{\infty} \sum_{m=0}^{\infty}P_l(\tau)P_n^1(\tau) P_m^2(\tau) E_{l+1,n, m}\left[W^H\left(V_{n},U_m, r+\sum_{i=1}^{l+1}X_i, \tau\right)\right]\\
&~~~-\lambda_1 \sum_{l=0}^{\infty} \sum_{n=0}^{\infty} \sum_{m=0}^{\infty}P_l(\tau) P_n^1(\tau) P_m^2(\tau) E_{n+1, m}\left[W^H\left(V_{n+1}, U_m r, \tau\right)\right]\\
&~~~-\lambda_2\sum_{l=0}^{\infty} \sum_{n=0}^{\infty} \sum_{m=0}^{\infty}P_l(\tau) P_n^1(\tau) P_m^2(\tau) E_{n, m+1}\left[W^H\left(V_{n},U_{m+1}, r, \tau\right)\right]\\
&=-\lambda\int_{0}^{\infty}\left(G\left(S_1,S_2, r+x, \tau\right)-G\left(S_1,S_2, r, \tau\right)\right) f_{r}(x)dx\\
&~~~-\lambda_1 \int_{0}^{\infty}\left(G\left(S_1y,S_2, r, \tau\right)-G\left(S_1,S_2, r, \tau\right)\right) f_{Y_1}(y)dy\\
&~~~-\lambda_2 \int_{0}^{\infty}\left(G\left(S_1,S_2y, r, \tau\right)-G\left(S_1, S_2, r, \tau\right)\right) f_{Y_2}(y)dy.
\end{aligned}
$$
The desired result follows.\qed

\section{Data simulation}

\setcounter{equation}{0}

\subsection{Series representation for $W$}

In this section, we give a series representation for $W\left(S, r, \tau\right)$ under model (\ref{3.7}), $\tau=T-t$.
For $W\left(S_1,S_2, r, \tau\right)$, we just need to replace the dsitribution of $S$ by the joint distribution of $H(S_1,S_2)$.
Remember that,
\beqnn
W\left(S, r, \tau\right)=E^{Q^{\top}}\left[\max \left(S^c(T)-K, 0\right) \mid S^c(t), r^c(t)\right]|_{S^c(t)=S,r^c(t)=r}. 
\eeqnn
Let $Z=\ln S^c$, $p(z)$ denote the probability density function of $Z$. By the results in He and Zhu (2017), we can get the following expression of this formula
$$
W\left(S, r, \tau\right)=P_1-K P_2,
$$
where
$$
\begin{aligned}
&P_1=f(-i ; t, T, z,r)\left\{\frac{1}{2}+\frac{1}{\pi} \int_0^{+\infty} \operatorname{Real}\left[\frac{e^{-i \phi l n K} f\left(\phi-i ; t, T, z, r\right)}{j \phi f\left(-i ; t, T, z, r\right)}\right] \mathrm{d} \phi\right\}, \\
&P_2=\frac{1}{2}+\frac{1}{\pi} \int_0^{+\infty} \operatorname{Real}\left[\frac{e^{-j\phi \ln K} f\left(\phi ; t, T, z, r\right)}{j \phi}\right] \mathrm{d} \phi,
\end{aligned}
$$
with 
$f\left(\phi ; t,T, z, r\right)=E^{Q^{\top}}\left[e^{i \phi Z(T)} \mid Z(t), r^c(t)\right]\big|_{ Z(t)=z,r^c(t)=r}
$ the characteristic function of $Z$. It is clear that in order
to obtain an explicit pricing process, we need to derive an analytical expression for the characteristic function of $Z$ under the forward measure
$Q^{\top}$, which results in the following theorem.

\btheorem\label{t4.4} 
If the underlying price follows the dynamics in model (\ref{3.7}), then the characteristic function has the following form,
\begin{equation}\label{5.1}
f\left(\phi ; t,T, z, r\right)=e^{B(\phi, \tau)+D(\phi, \tau) r+i \phi z} 
\end{equation}
where
$$
\begin{aligned}
&B(\phi ; \tau)= Ka\int_0^\tau D(\phi ; t) d t+\lambda\int_0^\tau D(\phi ; t)\int_{0}^{+\infty}x\left(e^{G(t, T) x}-1\right)f_r(x) dxdt+\frac{1}{2} i\phi \sigma^2 \tau-\frac{1}{2} \phi^2 \sigma^2 \tau  \\
&D(\phi ; \tau)=\frac{-2 \sum_{n=0}^{+\infty}(n+1) \hat{a}_{n+1} \tau^n}{\sigma_r^2 \sum_{n=0}^{+\infty} \hat{a}_n \tau^n},
\end{aligned}$$
with
$$
\begin{aligned}
&\hat{a}_{n+2}=-\frac{\hat{I}}{2 m(n+1)(n+2)}, n \geq 0, \quad \hat{a}_0=1, \quad \hat{a}_1=0;\\
& c_j=\frac{m^j}{j !},\,j\geq 1,  \quad  m=\sqrt{k^2+\sigma_r^2};\\
%&G(t, T)=-\frac{2\left(e^{m t}-1\right)}{2 m+(k+m)\left(e^{m t}-1\right)} \\
&\hat{I}=2 k m(n+1) \hat{a}_{n+1}+j \phi \sigma_r^2 m \hat{a}_n+(k+m) \sum_{j=1}^n(n+2-j)(n+1-j)c_j \hat{a}_{n+2-j} \\
&~~+\left(k^2+k m+2 \sigma_r^2\right) \sum_{j=1}^n(n+1-j) c_j \hat{a}_{n+1-j}+\frac{1}{2} i \phi \sigma_r^2(k+m) \sum_{j=1}^n c_j \hat{a}_{n-j}.
\end{aligned}
$$
\etheorem

The proof is similar to Theorem 2 in He and Zhu (2017), we omit it here. Similarly, we can get that if
\begin{equation}\label{5.2}
\tau \leq \frac{1}{m} \sqrt{\left[\ln \left(\frac{m-k}{m+k}\right)\right]^2+\pi^2},   m=\sqrt{k^2+\sigma_r^2},
\end{equation}
the series $\sum_{n=0}^{+\infty} \hat{a}_n \tau^n$ is always convergent. In the next section, matlab will be used for data simulation to test the convergence rate of its series solution.

\subsection{Data simulation}

In this section, the speed of convergence of the pricing formula is test with the parameter chosen under condition (\ref{5.2}), where $\sigma_r=\sigma=0.05$, $k=2$, $a=0.05$, $r_0=0.03$, $S_0=110$, $K=100$, $\lambda_1=1$, $\lambda_2=1$ and $X$ follows an exponential distribution with parameter $1000$. $Y\equiv 1.01$ a fixed jump amplitude.

\begin{figure}[H]
	\centering
	\begin{minipage}{0.49\linewidth}
		\centering
		\includegraphics[width=0.9\linewidth]{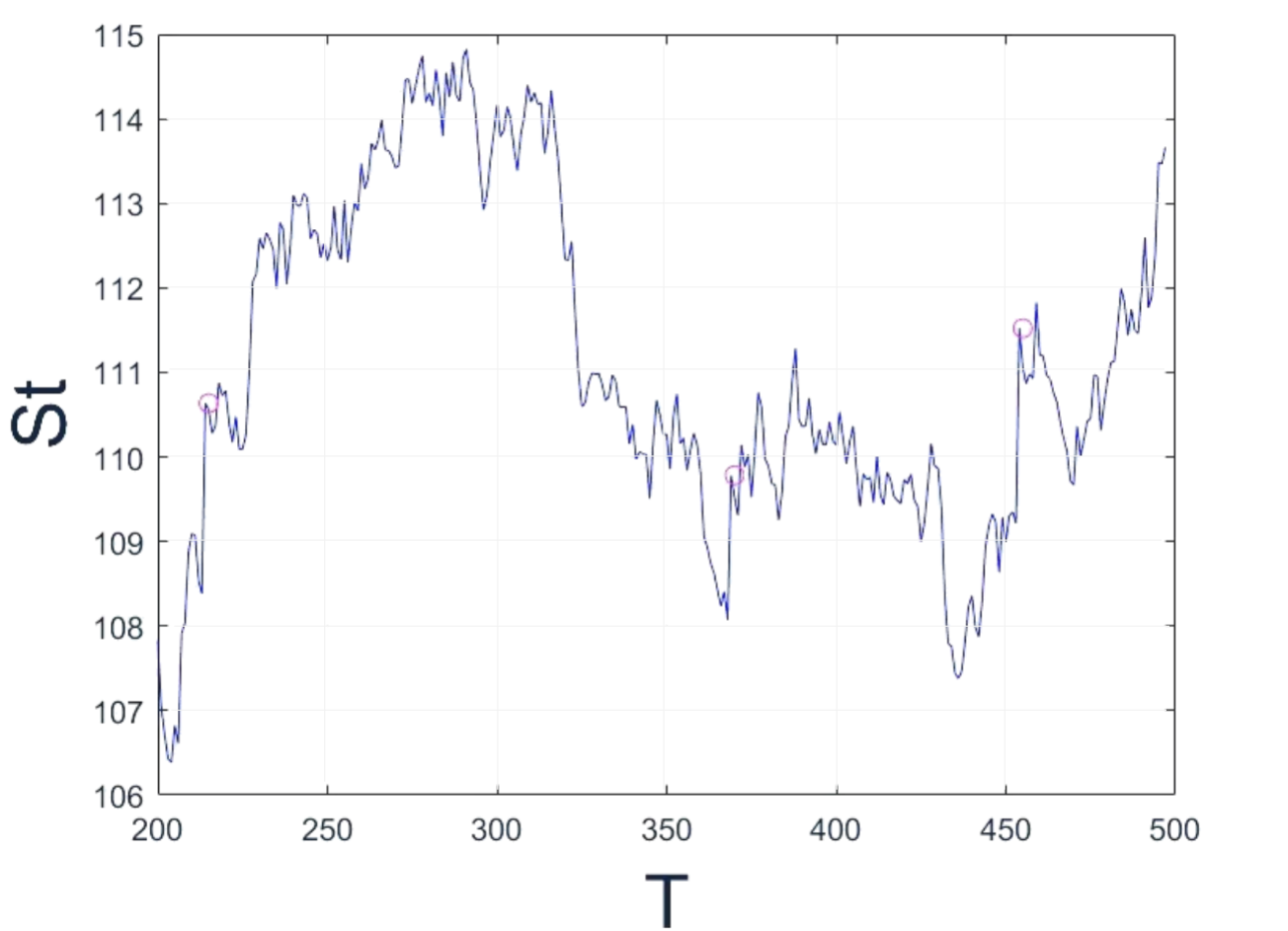}
		\caption{The path of $S(t)$}
		\label{chutian1}%文中引用该图片代号
	\end{minipage}
	\begin{minipage}{0.49\linewidth}
		\centering
		\includegraphics[width=0.9\linewidth]{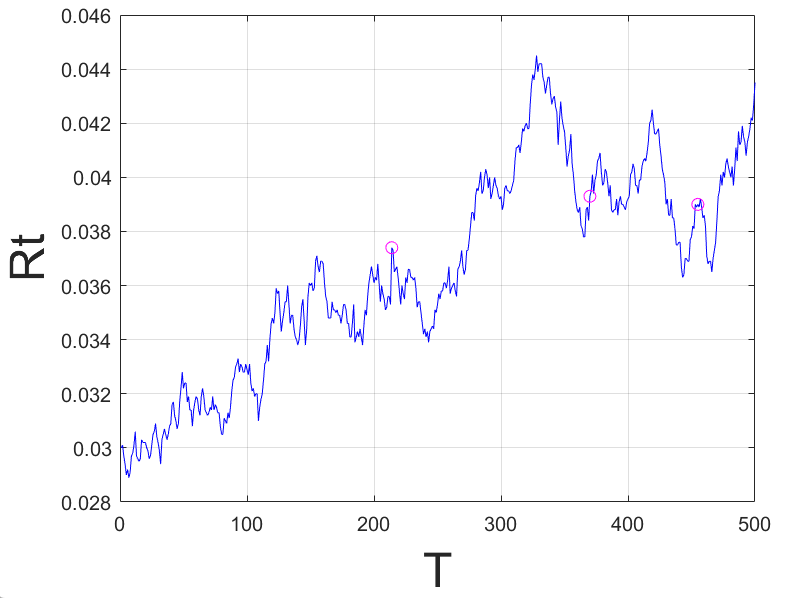}
		\caption{The path of $r(t)$}
		\label{chutian2}%文中引用该图片代号
	\end{minipage}
\end{figure}

\begin{figure}[H]
	\begin{minipage}{0.49\linewidth}
		\centering
		\includegraphics[width=0.9\linewidth]{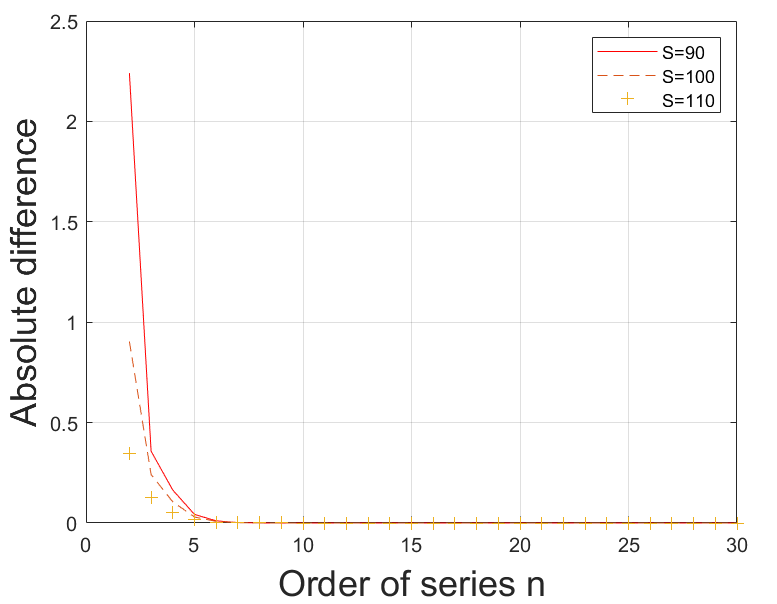}
		\caption{Absolute difference of $W\left(S,r,\tau\right)$ between $n+1$-term and $n$-term}
		\label{chutian3}%文中引用该图片代号
	\end{minipage}
	\begin{minipage}{0.49\linewidth}
		\centering
		\includegraphics[width=0.9\linewidth]{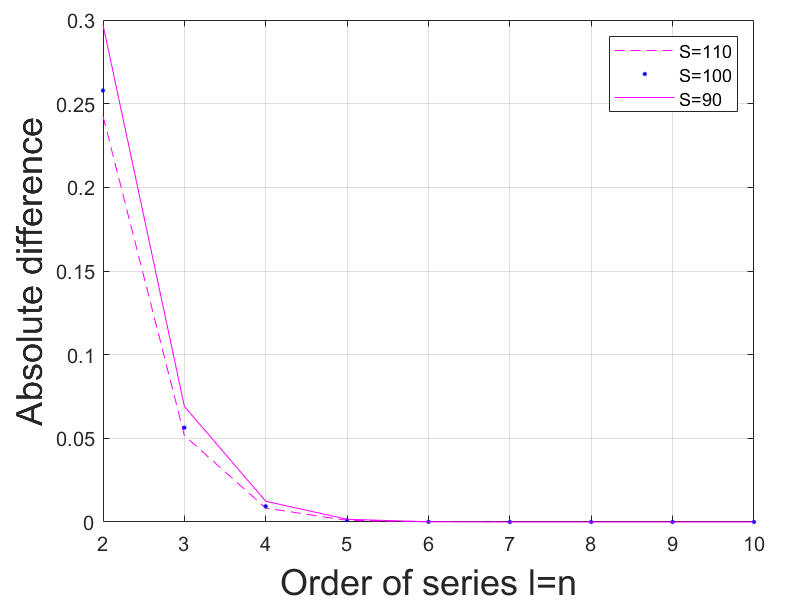}
		\caption{Absolute difference of $F(S, r, \tau)$ between $(l+1,n+1)$-term and $(l,n)$-term}
		\label{chutian4}%文中引用该图片代号
	\end{minipage}
\end{figure}

More than above, the absolute difference of $W\left(S,r,\tau\right)$ between $10$-term and $11$-term is of order $10^{-4}$. 
When $l=n=7$, the absolute difference of $F(S, r, \tau)$ between $(l+1,n+1)$-term and $(l,n)$-term is of order $10^{-4}$,
this can certainly show that the $(7,7)$-term price can be regarded as the converged option price for the double Lévy jump model.

\section{Conclusion}

In this paper, we adopt the hybrid model of double Lévy jump, including the underlying asset price  with Lévy jump and the CIR interest rate with Lévy jump. The pricing formula for European option in the form of series is given, and the numerical simulation by Matlab proves that our solution indeed converges very fast.

\bigskip

\textbf{Acknowledgement}. I would like to give my sincere thanks to the referee for her or his
valuable comments and suggestions.


\bigskip


\noindent
\begin{thebibliography}{99}
	
\bibitem{AJ92} Aim,K.I. and Jarrow, R.A.(1992). Pricing options on risky assets in a stochastic interest rate economy. \emph{Mathematical Finance} {\bf 4}, 217-237.
	
\bibitem{BS73} Black, F. and  Scholes, M. (1973). The pricing of options and corporate liabilities, \emph{Journal of Political Economy}. {\bf 81}, 637-659.

\bibitem{DF06} Brigo, D. and Mercurio, F.(2006). Interest Rate Models Theory and Practice[M]. Springer, Berlin, Heidelberg, 27-38.

\bibitem{CIR85} Cox, J.C., Ingersoll, J.E. and Ross S.A. (1985). A theory of the term structure of interest rates. {\emph Econometrica}. {\bf 53}, 385–406.

\bibitem{D03} Duffie, D., Filipovi$\acute{c}$, D. and Schachermayer, W. (2003). Affine processes and applications in
finance. \emph{Ann. Appl. Probab.} {\bf13(3)}, 984–1053.


\bibitem{HZ2017} He, X-J and Zhu, S-P. (2017). A closed-form pricing formula for European options under the Heston model with stochastic interest rate.\emph{Journal of Computational and Applied Mathematics}. {\bf 335}, 323-333.


\bibitem{HJM92} Heath, D., Jarrow, R. and Morton, A.(1992). Bond pricing and the term structure of interest rate: a new methodology for contingent claims Valuation,
\emph{Econometrica}. {\bf 60}, 77-105.

\bibitem{HW10} Heidari M. and Wu, L.(2010). Market anticipation of Fed policy changes and the term structure of interest rates. \emph{Review of Finance}. {\bf 14}, 313–42.

\bibitem{H93}  Heston, S.L. (1993). A closed-form solution for options with stochastic volatility with applications to bond and currency options.
\emph{Rev. Financ. Stud.} {\bf 6(2)}, 327–343.

\bibitem{HA90} Hull, J. and White, A. (1990). Pricing interest-rate derivative securities. \emph{Rev. Financ. Stud}. {\bf 3}, 1573–92.

\bibitem{J04} Johannes, M. (2004). The statistical and economic role of jumps in continuous-time interest rate models. \emph{Journal of Finance}. 227–60.

\bibitem{Ly99} Lin B-H, Yeh Shih-Kuo. (1999). Jump-diffusion interest rate process: an empirical examination. \emph{Journal of Business Finance $\&$ Accounting}. {\bf 26}, 967–95.

\bibitem{M76}  Merton, R.C. (1976). Option pricing when the underlying stock returns are discontinuous. \emph{Journal of Financial Economics}. {\bf 3}, 125-144.

\bibitem{PS13} Sattayatham P. and Pinkham. S. (2013). Option pricing for a stochastic volatility Lévy model with stochastic interest rates. \emph{Journal of the Korean Statistical Society} {\bf 42}, 25-36.

%\bibitem{ref7} Niu, L.Q. (2008) Some stability results of optimal investment in a simple Levy market \emph{Insurance: Mathematics and Economics}.  {\bf 42}, 445–452.

%\bibitem{S15} Steven E.Shreve. Stochastic calculus for finance[M]. Springer, New York, 2015.

\bibitem{S99} Sato, K. (1999). L$\acute{e}$vy process and Infinitely Divisible Distributions. Cambridge University Press.

\bibitem{V77} Vasicek, O. (1977). An equilibrium characterization of the term structure. \emph{J Financial Econ}. {\bf 5}, 177–88.


%\bibitem{ref9} Bruno Biais, Thomas BjörkJakša Cvitanić Nicole El Karoui,Elyés Jouini,Jean Charles Rochet,Wolfgang J. Runggaldier.Financial
%Mathematics[M].Springer,Berlin,Heidelberg,2006: $73 .$

%\bibitem{ref10} Corcuera JM.Nualart D.Schoutens W.Completion of a Lévy market by powerjump assets[J].Finance \& Stochastics, 2005, $9(1): 109-127 .$




\end{thebibliography}
\end{document}